\documentclass[%
 aip,
 amsmath,amssymb,
 reprint,%
]{revtex4-1}
\usepackage{graphicx}
\usepackage{dcolumn}
\usepackage{bm}

\usepackage[utf8]{inputenc}
\usepackage{braket}
\usepackage{graphicx}
\usepackage{amsmath}
\usepackage{amsfonts}
\usepackage{float}
\usepackage{amssymb}
\usepackage{bibunits}
\usepackage{epsfig}
\usepackage{epstopdf}
\DeclareGraphicsExtensions{.pdf,.eps,.png,.jpg,.mps}
\usepackage[pdftex]{color}
\usepackage{amsmath,graphicx,amssymb,braket,xcolor,subfigure,upgreek}
\usepackage[colorlinks, linkcolor=blue, citecolor=blue, urlcolor=blue, breaklinks=true]{hyperref}
\usepackage{microtype}
\usepackage{mathtools}
\usepackage{bbm}
\usepackage{color}

\newcommand{\red}[1]{\textcolor{black}{#1}}

\DeclareMathAlphabet{\mathcal}{OMS}{cmsy}{m}{n}

\newcommand{\sdag}{\sigma^{\dagger}}
\newcommand{\s}{\sigma}

\newcommand{\siginp}{\sigma_{\text{in}}}
\newcommand{\siginpd}{\sigma_{\text{in}}^{\dagger}}

\newcommand{\sir}{S_{\text{IR}}}
\newcommand{\BesselJ}{\mathcal{J}}

\begin{document}

\preprint{AIP/123-QED}

\title{Floquet engineering of molecular dynamics via infrared coupling}

\author{Michael Reitz}%
\email{Michael.Reitz@mpl.mpg.de}
\affiliation{
Max Planck Institute for the Science of Light, Staudtstra{\ss}e 2,
D-91058 Erlangen, Germany 
}%
\affiliation{
Department of Physics, University of Erlangen-Nuremberg, Staudtstra{\ss}e 7,
D-91058 Erlangen, Germany 
}%

\author{Claudiu Genes}
 \email{Claudiu.Genes@mpl.mpg.de}
\affiliation{
Max Planck Institute for the Science of Light, Staudtstra{\ss}e 2,
D-91058 Erlangen, Germany 
}%
\affiliation{
Department of Physics, University of Erlangen-Nuremberg, Staudtstra{\ss}e 7,
D-91058 Erlangen, Germany 
}%

\date{\today}

\begin{abstract}
We discuss Floquet engineering of dissipative molecular systems through periodic driving of an infrared-active vibrational transition, either directly or via a cavity mode. Following a polaron quantum Langevin equations approach, we derive correlation functions and stationary quantities showing strongly modified optical response of the infrared-dressed molecule. The coherent excitation of molecular vibrational modes, in combination with the modulation of electronic degrees of freedom due to vibronic coupling can lead to  both enhanced vibronic coherence as well as control over vibrational sideband amplitudes. The additional coupling to an infrared cavity allows for the controlled suppression of undesired sidebands, an effect stemming from the Purcell enhancement of vibrational relaxation rates.
\end{abstract}

\maketitle

\begin{figure}[b]
\includegraphics[width=0.94\columnwidth]{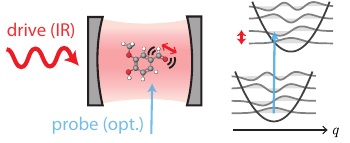}
\caption{\emph{Infrared-dressed molecule}. An infrared-active molecule is vibrationally dressed by the interaction with a quantized cavity field (coded in red). A weak optical probe field (coded in blue) can then detect the changes of the externally controlled vibronic dynamics by recording the absorption spectrum. The simultaneous coherent absorption of optical and infrared photons can in consequence modify the characteristics of the transition from the ground to the excited potential landscapes.}
\label{fig1}
\end{figure}

\section{\label{sec:level1}Introduction}
Over the last years,  coherent manipulation of quantum systems by means of (strong) periodic driving has been used to enable controlled engineering of matter states, from the single to the many-body level \cite{goldman2014periodically,oka2019floquet}. In particular, recently, the coherent excitation of phononic modes has been shown as a possibility to modify the properties of materials by steering them into non-equilibrium states. This includes dynamical control of electron-phonon coupling \cite{chenorbital2018, pomarico2017enhanced} and superconductivity \cite{buzzi2020photomolecular, babadi2017theory, murakami2017nonequilibrium}, topological phase transitions \cite{Huebener2018phonon}, and the control of real or synthetic magnetic fields \cite{Shin2018phonon, forst2015spatially, nova2017effective, schwennicke2020optical}.
Some of the experimental techniques rely on the excitation of phonon modes through acoustic modulation. Alternatively, the coupling to infrared-active vibrational modes provides another way to drive and modulate vibrational degrees of freedom. In confined electromagnetic environments such as cavities and plasmonic nanostructures, the collective vibrational strong coupling of molecular ensembles to infrared modes leads to the occurence of vibrational polaritons and therefore modifies the vibrational dynamics \cite{Shalabney2015, Long2015coherent, pino2015quantum, dunkelberger2016modified, xiang2018twodimensional}. This has been e.g.~shown to reveal enhanced Raman scattering \cite{shalabeney2015enhanced} or changes of chemical reaction kinetics \cite{Thomas2019tilting, thomas2020ground, lathercavity2019, galego2019cavity}. The coupling of vibrational modes to plasmonic resonators has also been used to increase spectroscopic sensitivity through surface-enhanced infrared absorption \cite{adato2009ultra, neubrech2008resonant}.\\
\indent While \textit{Floquet engineering} (i.e.,~the dressing of matter with oscillating fields) has been extensively studied at the level of unitary, pure Hamiltonian evolution, such as realized in well-isolated systems like cold atoms in optical lattices \cite{ eckardt2017colloquium, goldman2015periodically, schweizer2019floquet}, its application to dissipative, open quantum systems is much less explored (for exceptions see  e.g.~Refs.~\cite{restrepo2016driven,schnell2020is, tatsuhiko2016effective, haddadfarshi2015completely}). Although isolated molecules would, in principle, represent almost coherent quantum optical systems, in practice, their coupling to condensed matter environments like crystals or solvents leads to strong decoherence.  The main decoherence mechanisms are a temperature-dependent electronic dephasing stemming from higher-order electron-phonon coupling \cite{muljarov2004dephasing, clear2020phonon} as well as vibrational relaxation due to coupling of the localized molecular vibrational modes (\textit{vibrons}) to the surrounding thermal bath of phonons \cite{reitz2020molecule, hill1988vibrational}.\\
\indent We consider here the case of \textit{infrared-dressed} molecules, where the vibronic coupling is modified by periodic driving of vibrational transitions achieved either directly by the application of a laser, or indirectly by coupling to a quantized driven electromagnetic mode of a cavity (see Fig.~\ref{fig1}). The immediate effect is the modification of the standard Franck-Condon physics as the transition matrix elements for optical absorption from electronic ground to excited state are largely impacted by the coherent infrared drive. The main mechanism is the competition between coherent processes, such as the simultaneous absorption of probe and drive photons and the standard path of direct optical photon absorption. For a proper treatment of dissipative dynamics we follow a quantum Langevin equations approach for the \textit{polaron operator} (i.e.,~a vibrationally dressed electronic dipole operator). This allows the inclusion of all coherent and incoherent effects at the level of the equations of motion for system operators. By formally integrating the dynamics, we analytically estimate two-time correlations and derive stationary quantities aimed at qualitatively and quantitatively characterizing the modified response of the dressed molecule to weak probe optical fields.\\
\indent A remarkable result is the strong modification of vibronic sidebands which, under external infrared driving, can exhibit both enhanced amplitudes and reduced linewidths. The improved lifetime stems from a competition between vibrational relaxation and gain from coherent optical and vibrational processes. We show that the effect can lead to improved vibronic coherence. The embedding within an infrared, cavity-confined electromagnetic mode brings an enhancement of the interaction by a factor proportional to the cavity finesse. Moreover, an infrared cooling effect can increase the vibrational relaxation in a wide bandwidth, thus effectively reducing undesired scattering. This can be understood as a cavity Purcell effect at the level of the vibrational relaxation damping rate.\\
\section{\label{sec:level2}Theory and methods}
We consider a molecule with electronic ground $\ket{g}$ and excited $\ket{e}$ states, separated by an energy difference $\omega_0$ (we set $\hbar=1$) and define the Pauli raising ${\sigma}^\dagger=\ket{e}\bra{g}$ and lowering ${\sigma}=\ket{g}\bra{e}$ operators. \red{We only consider the harmonic approximation of the ground and excited potential landscapes along a given single nuclear coordinate as illustrated in Fig.~\ref{fig1}. We proceed by providing a derivation of the vibronic coupling (between electron and vibrational mode) in the harmonic approximation.} We will furthermore assume identical frequencies $\nu$ of the potential energy surfaces for both electronic orbitals. We then describe the open system dynamics including both coherent and incoherent effects in terms of a master equation. We then analyze the modification of the standard Franck-Condon physics, stemming from the modulation of the system's energy levels induced by the infrared driving. Finally, we introduce the transition from the master equation to quantum Langevin equations, approach which allows for the derivation of analytical results.
\begin{figure*}[t]
\includegraphics[width=1.8\columnwidth]{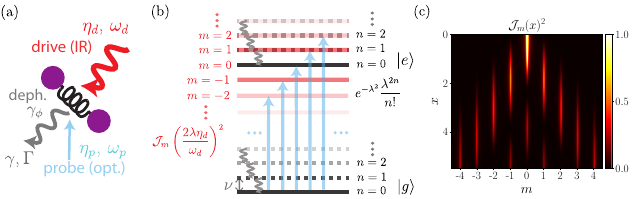}
\caption{\emph{Floquet quasienergies \& transitions}. (a) A single nuclear coordinate \red{that can be imagined as the vibration of a diatomic molecule} is driven with amplitude $\eta_d$ and frequency $\omega_d$. The optical transition is probed by a weak field with amplitude $\eta_p$ and frequency $\omega_p$. The molecule is embedded in a dephasive environment leading to a temperature-dependent broadening with rate $\gamma_\phi$. Additional losses are radiative ($\gamma$) and vibrational relaxation ($\Gamma$). (b) Quasienergy scheme of an infrared-dressed molecule in the case of $\omega_d=\nu$. Additionally to the usual vibronic sidebands (grey dashed lines) with transition strengths given by a Poissonian distribution $e^{-\lambda^2}\lambda^{2n}/n!$, one obtains a set of driving-induced quasienergies at $\omega_0+m\omega_d$ with transition strengths  proportional to Bessel functions squared $\BesselJ_m\left({2\lambda\eta_d}/\omega_d\right)^2$. Vibrational relaxation is indicated by wiggly lines. (c)  Bessel functions $\BesselJ_m(x)^2$ determining the intensities of the driving-induced sidebands. }
\label{fig2}
\end{figure*}
\subsection{Hamiltonian description \& master equation approach}
\red{Assuming equilibrium positions $R_g$ and $R_e$ for the potential surfaces of electronic ground and excited states, one can write the total molecular Hamiltonian describing both electronic and vibrational dynamics as
\begin{align}
\label{molham}
\mathcal{H}=&\left[\omega_0+\frac{\hat{P}^2}{2\mu}+\frac{1}{2}\mu\nu^2\left(\hat{R}-R_e\right)^2\right]\sigma^\dagger\sigma \\\nonumber
&+ \left[\frac{\hat{P}^2}{2\mu}+\frac{1}{2}\mu\nu^2\left(\hat{R}-R_g\right)^2\right]\sigma\sigma^\dagger,
\end{align}
where $\mu$ is the reduced mass of the vibrational mode. The kinetic and potential energies are written in terms of the position $\hat{Q}$ and momentum operator $\hat{P}$ describing the nuclear coordinate under consideration, with commutation $[\hat{Q},\hat{P}]=i$. While in reality anharmonic effects come into play already at the level of a few excitations (especially in the case of small molecules), the harmonic expansion is a good approximation, especially for low temperatures and weak external driving which sees less than one vibrational excitation reached in steady state. Introducing oscillations around the equilibria $\hat{Q}=\hat{R}-R_{\text{g}}$ and subsequently $\hat{R}-R_{\text{e}}=\hat{Q}+R_{\text{g}}-R_{\text{e}}=:\hat{Q}-R_{\text{ge}}$ we obtain
\begin{equation}
\mathcal{H}=\frac{\hat{P}^2}{2\mu}+\frac{1}{2}\mu\nu^2 \hat{Q}^2 +\omega_0\sigma^\dagger\sigma-\mu\nu^2 \hat{Q} R_{\text{ge}}\sigma^\dagger\sigma+\frac{1}{2}\mu\nu^2 R_{\text{ge}}^2 \sigma^\dagger\sigma.
\end{equation}
 We can now rewrite the momentum and position operators in terms of bosonic operators $\hat{Q}=q_{\text{zpm}}(b^\dagger+b)$, $\hat{P}=ip_{\text{zpm}}(b^\dagger-b)$. The bosonic operators satisfy the usual commutation relation $[{b},{b^\dagger}]=1$ and the zero-point motion displacement and momentum are defined as $q_{\text{zpm}}=1/\sqrt{2\mu\nu}$ and $p_{\text{zpm}}=\sqrt{\mu\nu/2}$. Reexpressing the terms above yields the Holstein Hamiltonian \cite{holstein1959study}}
\begin{align}
\label{holsteinhamiltonian}
\mathcal{H}=(\omega_0+\lambda^2\nu){\sigma}^\dagger{\sigma}+\nu {b}^\dagger{b}-\lambda\nu({b}^\dagger+{b}){\sigma}^\dagger{\sigma}.
\end{align}
The dimensionless vibronic coupling strength $\lambda$ is given by $\lambda=\mu\nu R_{\text{ge}}q_{\text{zpm}}$ ($\lambda^2$ is called the Huang-Rhys factor and is typically on the order of $\sim 0.01-1$).\\
\indent An external infrared field $\mathcal{E}(q,t)$ can couple to the molecular vibrations via the displacement-dependent molecular dipole operator ${\mu}(q)$. In the dipolar approximation $\mathcal{E}(q,t)\approx \mathcal{E}(t)$, the Hamiltonian describing the coupling is given by ${\mathcal{H}}_{\text{IR}}(t)=-{\mu}(q)\cdot{\mathcal{E}}(t)$. A first order Taylor expansion around the equilibrium positions of the nuclei (subscript ``0''); $\braket{{\mu}(q)}=\braket{{\mu}}_0+(\partial\braket{{\mu}}/\partial q)_0\,{q}$ then gives rise to a coupling between the nuclear displacement and the electric field. To engineer an \textit{infrared-dressed} molecule, we consider a strong classical drive of the molecular vibrational mode with amplitude $\eta_d$ and frequency $\omega_d$ expressed in Hamiltonian form as
\begin{align}
\mathcal{H}_d=\eta_d\cos(\omega_d t)(b^\dagger+{b}).
\end{align}
Simultaneously, we consider a weak optical field probing the electronic transition at some variable laser frequency $\omega_p$ and small amplitude $\eta_p$ modelled as a standard drive Hamiltonian $\mathcal{H}_p=i\eta_p (\sigma^\dagger e^{-i\omega_p t}-\sigma  e^{i\omega_p t})$.

\indent Dissipative processes are included as standard Lindblad terms characterized by a collapse operator acting at a given loss rate. For example, radiative spontaneous emission of the electronic excited state can be cast in Lindblad form by $\mathcal{L}_\sigma^\gamma[\rho]=\gamma(2\sigma\rho\sigma^\dagger-\{\sigma^\dagger\sigma,\rho\})$ where $\rho$ is the density operator and the collapse operators $\sigma$ describes de-excitation of the electronic excited state at rate $\gamma$. Pure dephasing  of the electronic transition (\red{i.e., decay of the electronic coherences}) with rate $\gamma_\phi=\gamma_\phi(T)$ \red{due to quadratic electron-phonon coupling} can be included by the Lindblad term $\mathcal{L}_{\sigma^\dagger\sigma}^{\gamma_\phi}[\rho]$ with collapse operator $\sigma^\dagger\sigma$. This is a strongly temperature-dependent term ($\propto T^2$ at low temperatures) which is close to zero at cryogenic temperatures ($T \lesssim 4\,\text{K}$) where experiments observe lifetime-limited zero-phonon line (ZPL) transitions \cite{pazzagli2018self,wang2019turning}. While the effect of temperature is included as a dephasing of the electronic transition, thermal occupation of the vibrational modes can commonly be neglected even at elevated temperatures since molecular vibrations typically lie at high terahertz frequencies (i.e.,~we assume $k_{\text{B}} T/(\hbar\nu)\ll1$). However, analytical treatments have shown methods to include thermal effects as non-zero occupancy of the vibrational mode \cite{reitz2020molecule}.  Finally, a master equation can be written to describe the open system dynamics
\begin{equation}
\dot{\rho}(t)=i[\rho(t),\mathcal{H}+\mathcal{H}_{{d}}+\mathcal{H}_{{p}}]+\mathcal{L}[\rho(t)],
\end{equation}
which  generally provides numerical support but less possibilities for analytical calculations.

\subsection{Floquet states and transition strengths}
The Holstein Hamiltonian can be diagonalized via a level-dependent polaron transformation $\mathcal{U}^\dagger=\ket{g}\bra{g}+\mathcal{D}^\dagger\ket{e}\bra{e}$ with the standard displacement operator $\mathcal{D}=e^{-i\sqrt{2}\lambda p}=e^{\lambda(b^\dagger-b)}$. In the polaron-displaced basis, the Holstein Hamiltonian becomes $\tilde{\mathcal{H}}=\mathcal{U}^\dagger \mathcal{H}\mathcal{U}=\omega_0{\sigma}^\dagger{\sigma}+\nu {b}^\dagger{b}$ and has simple eigenvectors $\ket{g;n}$ and $\ket{e;n}$. The eigenvectors in the bare, original basis can be found by inverting the polaron transformation $\ket{g;n}$ and $\mathcal{D}\ket{e;n}$. The polaron-transformed probe Hamiltonian is then expressed as $\tilde{\mathcal{H}}_p=i\eta_p (\sigma^\dagger \mathcal{D}^\dagger e^{-i\omega_p t}-\sigma \mathcal{D} e^{i\omega_p t})$. We can now look for selection rules applying to processes such as stimulated emission and absorption induced by the external optical drive. To this end we focus on absorption (as emission is similar) by assuming an initial state $\ket{g;0}$ in the displaced basis and asking for the probability of exciting the system to state $\ket{e;n}$. This is easily computed to lead to
\begin{equation}
\label{poissonian}
P_\text{abs}(n)=|\bra{e;n}\sigma^\dagger\mathcal{D}^\dagger\ket{g;0}|^2=e^{-\lambda^2}\frac{\lambda^{2n}}{n!},
\end{equation}
which is the expected Poissonian distribution leading to the Franck-Condon principle for molecular transitions. For dissipative radiative processes, we notice that the Lindblad collapse operator is also transformed to the polaron one $\sigma \mathcal{D}$ such that spontaneous emission follows the same Poissionian distribution in taking the electronic state from $\ket{e;0}$ to $\ket{g;n}$.\\
\indent The periodic driving can change this physics drastically. The drive Hamiltonian in the polaron picture becomes
\begin{align}
\tilde{\mathcal{H}}_d=\eta_d\cos(\omega_d t)(b^\dagger+{b}+2\lambda\sigma^\dagger\sigma).
\end{align}
The effect on the electronic transition is a periodic modulation of the excited state energy with $\tilde{\omega}_0(t)=\omega_0+2\lambda\eta_d\cos(\omega_d t)$. Let us now perform a transformation into an interaction picture with the following unitary operator $\mathcal{U}_{\text{int}}=e^{i\sigma^\dagger\sigma\int_0^t dt'\tilde{\omega}_0(t')+i\nu b^\dagger b t}$. Moreover we also expand the displacement operator $\mathcal{D}^\dagger$ in terms of annihilation and creation operators in order to reveal the fundamental processes involved in the absorption and emission of optical photons. We also introduce $\BesselJ_m(x)$ as Bessel functions of the first kind and the probe beam detuning $\Delta_p=\omega_p-\omega_0$. We can then find the following expression for the probe Hamiltonian in the displaced interaction picture
\begin{widetext}
\begin{align}
\label{hamiltonianinteraction}
\tilde{\mathcal{H}}_p^{\text{int}}=\eta_p e^{-\lambda^2/2}\left (\sigma^\dagger \sum_{k,\ell=0}^\infty \frac{(-\lambda)^k\lambda^\ell}{\ell!\,k!}\left(b^\dagger e^{i\nu t}\right)^k\left(b e^{-i\nu t}\right)^\ell\sum_{m=-\infty}^\infty \BesselJ_m\left(\frac{2\lambda\eta_d}{\omega_d}\right)e^{im\omega_d t}e^{-i\Delta_p t}-\text{h.c.}\right).
\end{align}
\end{widetext}
\indent Additionally, one is still left with a term $\tilde{\mathcal{H}}_d^{\text{int}}=\eta_d\cos(\omega_d t)(b^\dagger e^{i\nu t} + b e^{-i\nu t})$ which can drive transitions between vibrational states and becomes very important for $\omega_d\approx\nu$. Let us first understand the above expression on the simplified case of $\eta_d=0$ where we expect the usual selection rules for Franck-Condon physics. In this case, only the term with $m=0$ contributes and operator combinations of the kind $\sigma^\dagger (b^\dagger)^k b^\ell$ have time dependence $e^{i[\Delta_p-(k-\ell)\nu]t}$. Imposing resonance allows to find the transition strength for different processes. For example, the transition from $\ket{g;0}$ to $\ket{e;n}$ can only be achieved by $k=n$ and $\ell=0$. The resonance condition then asks that the laser frequency is $\omega_p=\omega_0+n\nu$ and the strength is proportional to $e^{-\lambda^2/2}\lambda^n/\sqrt{n!}$ as expected.\\
\indent For $\eta_d\neq 0$, the driving \red{induces} a set of additional resonances at $\omega_p=\omega_0+m\omega_d$ where $m=0,\pm 1, \pm 2,\dots$. In the vicinity of these resonances, the Hamiltonian in Eq.~(\ref{hamiltonianinteraction}) can be transformed back into a time-independent frame
\begin{align}
\tilde{\mathcal{H}}_m=&-(\Delta_p\!-\!m\omega_d)\sigma^\dagger\sigma+\nu b^\dagger b\\\nonumber
&+\!i\eta_p\!\left(\sigma^\dagger\mathcal{D}^\dagger \BesselJ_m\left(\!\frac{2\lambda\eta_d}{\omega_d}\!\right)\!-\!\text{h.c.}\!\right),
\end{align}
with the driving-induced Floquet quasienergies of the electronic excited state $\omega_0 + m \omega_d$.  In the Jablonski scheme in Fig~\ref{fig2}(b) we illustrate the energy levels and corresponding transitions  for resonant driving $\omega_d=\nu$. In this case, some of the Floquet states $\ket{e;m}$ become degenerate to the vibrational states $\ket{e;n}$ ($m\geq 0$). The selection rules for the modified transition probabilities $P_\text{abs}(n)$ are then given by $\ell=0$, $n=k+m$ and the transition probability can be summed up to
\begin{align}
P_\text{abs}(n)=&e^{-\lambda^2} \sum_{m=-\infty}^n \frac{\lambda^{2(n-m)}}{(n-m)!}\BesselJ_m\left(\frac{2\lambda\eta_d}{\omega_d}\right)^2.
\end{align}
In the Floquet picture, each of the above terms corresponds to the absorption of a photon from the ground state $\ket{g;0}$ to $\ket{e;n-m}$ ($m\in [-\infty,n]$) in combination with the simultaneous dressing by $|m|$ infrared photons. The term corresponding to $n=m$ in the above sum with probability $e^{-\lambda^2}\BesselJ_n(2\lambda\eta_d/\omega_d)^2$ describes a transition on the zero-phonon line (\red{or `zero-vibron line'}) in combination with the absorption of $n$ infrared photons. In the following sections we will show that this term can become dominant and can be used to modify the vibronic linewidths and coherences since the zero-phonon line is not affected by the vibrational relaxation $\Gamma$.

\subsection{Langevin equations of motion}
The aforementioned master equation can be mapped onto a stochastic Heisenberg-Langevin equation~\cite{gardiner2004quantum,gardiner1985input,reitz2019langevin}. Explicitly, for a Lindblad term $\mathcal{L}_c^{\gamma_c}[\rho]$ with collapse operator $c$ and decay rate $\gamma_c$, the equation of motion for any operator $\mathcal{O}$ becomes
\begin{align}
\dot{\mathcal{O}}&=i[H,\mathcal{O}]-[\mathcal{O},c^\dagger]\left\{\gamma_c c-\sqrt{2\gamma_c}c_{\text{in}}\right\}+\\\nonumber
&+\left\{\gamma_c c^\dagger-\sqrt{2\gamma_c}c_{\text{in}}^\dagger\right\}[\mathcal{O},c^\dagger],
\end{align}
where each decay process is associated with a zero-averaged stochastic input noise operator $c_{\text{in}}$ with delta correlations in time $\braket{c_{\text{in}}(t)c_{\text{in}}^\dagger(t')}=\delta(t-t')$. Additionally, the vibrational relaxation of the molecule with rate $\Gamma$ due to the coupling of the vibron to a (harmonic) bath of phonons is included as a Brownian noise dissipation model (for details see below) which cannot be expressed as a simple master equation in Lindblad form \cite{reitz2020molecule, hu1992quantum}. We note that for simplicity, for the numerical simulations presented in this work, we use a Lindblad decay model for the vibration with Lindblad term $\mathcal{L}_b^{\Gamma}[\rho]$ which becomes identical to the Brownian noise model in the limit of $\lambda^2\Gamma\ll\gamma$ \cite{reitz2019langevin}. We note that the vibron also undergoes spontaneous \red{IR}-emission due to its coupling to the electromagnetic vacuum which we will however neglect since this radiative emission is much slower than the vibrational relaxation \cite{metzger2019purcell}. \\
\indent In a frame rotating at the probe frequency $\omega_p$, under the assumption of weak probe fields $\eta_p\ll\gamma$ and therefore $\braket{\sigma^\dagger\sigma}\ll 1$, the simplified quantum Langevin equations of motion for the mechanical degrees of freedom ($\{q,p\}$)  as well as for the polaron operator $\tilde{\sigma}:=\sigma\mathcal{D}^\dagger$ are given by (see Appendix for derivation and full equations of motion)
\begin{subequations}
\label{eqnsfreespace}
\begin{align}
\dot{q}&=\nu p,\\
\dot{p}&=-2\Gamma p-\nu q -\sqrt{2}\eta_d \cos(\omega_d t)+\xi(t),\\
\dot{\tilde{\sigma}}&=-(\tilde{\gamma}-i\Delta_p)\tilde{\sigma}-2i\lambda\eta_d\cos(\omega_d t)\tilde{\sigma}+\tilde{\Sigma}_{\text{in}},
\label{sigmaformal}
\end{align}
\end{subequations}
where  $\tilde{\gamma}=\gamma+\gamma_\phi$. Here, we defined the input affecting the electronic transition as \
\begin{equation}
\tilde{\Sigma}_{\text{in}}=\left[\eta_p+\sqrt{2\gamma}{\sigma}_{\text{in}}+\sqrt{2\gamma_\phi}\left((\sigma^\dagger\sigma)_{\text{in}}\sigma-\sigma(\sigma^\dagger\sigma)_{\text{in}}\right)\right]\mathcal{D}^\dagger,
\end{equation}
with correlations of the zero-average electronic noise operators given by $\braket{{\sigma}_{\text{in}}(t){\sigma}_{\text{in}}^\dagger (t')}=\delta(t-t')$ and $\braket{(\sigma^\dagger\sigma)_{\text{in}}(t)(\sigma^\dagger\sigma)_{\text{in}}(t')}=\delta(t-t')$. From Eq.~(\ref{sigmaformal}) one can see that, as previously discussed, the driving of the vibrational mode also leads to an effective modulation of the electronic transition frequency $\tilde{\omega}_0=\omega_0+2\lambda\eta_d\cos(\omega_d t)$.  The correlations of the zero-average Brownian noise stochastic force ${\xi}(t)$ affecting the momentum of the vibrational mode are given by
\begin{align}
\braket{\xi(t)\xi(t')}=\frac{1}{2\pi}\int_{-\infty}^\infty d\omega e^{-i\omega(t-t')}S_{\text{th}}(\omega),
\end{align}
with the colored thermal spectrum $S_{\text{th}}(\omega)=2\Gamma\omega\left[\coth(\hbar\omega/(k_{\text{B}}T))+1\right]/\nu$. For $k_{\text{B}} T/(\hbar\nu)\ll1$, the thermal spectrum becomes asymmetric and can be approximated by $S_{\text{th}}(\omega)=4\Gamma\omega/\nu\theta(\omega)$ ($\theta(\omega)$ is the Heaviside function). To calculate correlations of the vibrational mode, one can generally proceed with a Fourier analysis and make use of the fact that the noise is always $\delta$-correlated in the frequency domain $\braket{\xi(\omega)\xi(\omega')}=S_{\text{th}}(\omega)\delta(\omega+\omega')$.
After formal integration of Eq.~(\ref{sigmaformal}), one can obtain a solution for the electronic coherence $\braket{\sigma(t)}$.
 The expression for the population of the electronic excited state $\braket{\sigma^\dagger\sigma}$ in the long-time limit is then given by
\begin{align}
\braket{\sigma^\dagger\sigma}(t)=2\eta_p\int_{-\infty}^t dt' e^{-2\gamma(t-t')}\Re\braket{\sigma(t')}.
\end{align}
In the following section we show that this expression becomes stationary under some assumptions and can then be interpreted as the steady-state absorption spectrum.\\
We start by describing how the driving of the vibrations modifies the molecular absorption spectrum. Driving close to resonance $\omega_d\approx\nu$ will steer the vibron into a coherent state and will lead to a non-zero steady state occupation of the vibrational mode $|\beta|^2:=\braket{b^\dagger b}=(\eta_d/2)^2/\left(\Gamma^2+\Delta_d^2\right)$ with the detuning $\Delta_d=\omega_d-\nu$ and $\beta=(\eta_d/2)/\left(\Gamma-i\Delta_d\right)$. For the calculation of the absorption spectrum, one has to evaluate the displacement correlation function $\braket{\mathcal{D}(t)\mathcal{D}^\dagger (t')}$ which is non-stationary due to the driving of the vibrational mode. The calculation of $\braket{\mathcal{D}(t)\mathcal{D}^\dagger (t')}$ requires the momentum correlation function $\braket{p(t)p(t')}$ (full expression specified in Appendix) whose stationary part for a given time ordering $t\geq t'$ equates to
\begin{align}
\braket{p(t)p(t')}_{\text{st}}=\frac{e^{-(\Gamma+i\nu)(t-t')}\!}{2}+|\beta|^2\cos(\omega_d (t-t')),
\end{align}
where the first part describes the free evolution of the vibrational mode and the second part describes the modification due to the driving. The displacement correlation function (see Appendix for derivation) can then be evaluated as
\begin{align}
\label{corr}
\braket{\mathcal{D}(t)\mathcal{D}^\dagger (t')}\!=\!e^{-\lambda^2}e^{\lambda^2\mathrm{exp}\left[-(\Gamma+i\nu)(t-t')\right]}e^{-i\sqrt{2}\lambda\braket{p(t)-p(t')}},
\end{align}
with $\braket{p(t)}=\left(\beta^* e^{i\omega_d t}+\beta e^{-i\omega_d t}\right)/\sqrt{2}$ the time-dependent momentum expectation value.

\begin{figure*}[t]
\includegraphics[width=2.03\columnwidth]{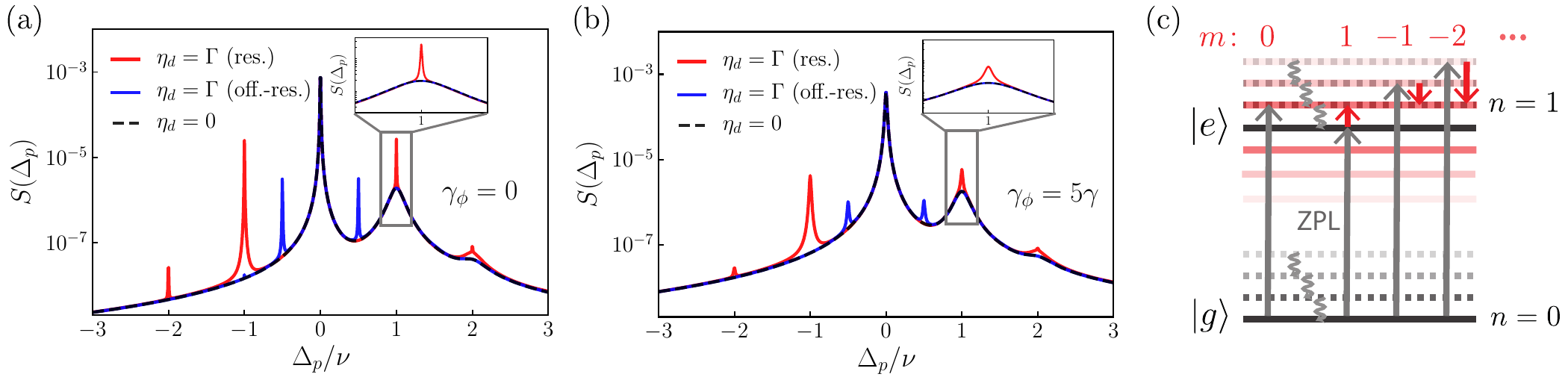}
\caption{\emph{Absorption spectrum \& sideband control}. (a) Modification of molecular absorption spectrum (logarithmic scale) by means of off-resonant driving (blue, $\omega_d=0.5\nu$, $|\beta|^2\approx 0$), and resonant driving (red, $\omega_d=\nu$, $|\beta|^2=1/4$) in the low-temperature (zero dephasing) regime for $\lambda=0.2$. The inset shows a zoom-in on the first vibronic sideband at $\Delta_p=\nu$. Other parameters: $\Gamma=0.1\nu=50\gamma$, $\gamma=5\cdot10^{-3}\nu$, $\eta_p=10^{-2}\gamma$. (b) Absorption spectrum for the same parameters as in (a) but finite dephasing $\gamma_\phi=5\gamma$. (c) Floquet contributions to first vibronic sideband $\Delta_p=\nu$ in the case of resonant driving $\omega_d=\nu$: pure vibronic transition (left arrow, $m=0$) and transitions due to dressing with infrared field ($m\neq 0$). For $\lambda^2\ll1$, the transitions with negative $m$ and $n\geq 2$ are extremely unlikely due to small Franck-Condon overlap. Additional transitions can stem from non-zero vibrational occupation ($\ell\neq 0$).}
\label{fig3}
\end{figure*}

\section{\label{sec:level3}Results}
We proceed by analysing the response of an infrared driven molecule to a weak optical probe and show the possibility of controlling the linewidth of vibrational sidebands. We then analyze the increase in coherence for the dipole transition stemming from the reduction in vibrational relaxation rates. Finally, we show that indirect driving through an infrared cavity can exhibit narrow isolated sidebands via the cavity Purcell effect of vibrational relaxation rates.
\subsection{Sideband control via Floquet engineering}

While the ground-excited coherence $\braket{\sigma(t)}$ is generally not stationary, under the assumption of $\gamma\ll\omega_d$, the excited state population (i.e.,~the absorption spectrum) $\braket{\sigma^\dagger\sigma}=:S(\Delta_p)$ becomes time-independent and expresses as
\begin{align}
\label{modabsorption}
S(\Delta_p)&\!=\!\frac{\eta_p^2}{\gamma}\!\sum_{\substack{n=0\\ m,\ell=-\infty}}^\infty\!\frac{s_n^\lambda(\tilde{\gamma}\!+\!n\Gamma)\mathcal{J}_m\!\left(\frac{2\lambda\eta_d}{\omega_d}\right)^2\BesselJ_\ell\left(2\lambda|\beta|\right)^2}{(\tilde{\gamma}+n\Gamma)^2+(\Delta_p\!-\!n\nu-\!(m\!+\!\ell)\omega_d)^2},
\end{align}
In addition to the usual vibronic sidebands weighted by a Poissonian distribution $s_n^\lambda=e^{-\lambda^2}\lambda^{2n}/n!$ at frequencies $n\nu$ ($n\in\mathbb{N}_0$) and linewidths $\tilde{\gamma}+n\Gamma$, the driving induces a series of Lorentzian sidebands weighted by Bessel functions of the first kind $\BesselJ_m\left(2\lambda\eta_d/\omega_d\right)^2$ at multiples of the driving frequency $m\omega_d$ ($m\in\mathbb{Z}$)  and a series of sidebands weighted by Bessel functions $\BesselJ_\ell(2\lambda|\beta|)^2$, also at multiples of the driving frequency $\ell\omega_d$ ($\ell\in\mathbb{Z}$). While the former stems, as previously discussed, from the shaking of the electronic transition and is discussed in more detail below, the latter accounts for the finite occupation of the vibrational mode and comes into play in case of resonant driving of the vibron, i.e., if $\omega_d\approx\nu$. \red{We also note that the result in Eq.~(\ref{modabsorption}) fulfills the standard sum rules for optical transitions, i.e., the oscillator strengths sum up to unity
\begin{align}
e^{-\lambda^2}\sum_{n=0}^\infty\frac{\lambda^{2n}}{n!}\sum_{m,\ell=-\infty}^\infty \mathcal{J}_m\left(\frac{2\lambda\eta_d}{\omega_d}\right)^2\mathcal{J}_\ell\left(2\lambda|\beta|\right)^2=1.
\end{align}}
  In Fig.~\ref{fig3}(a) we plot the absorption spectrum $S(\Delta_p)$ in the limit of $\tilde{\gamma}\ll\Gamma$ and zero dephasing for both resonant and off-resonant driving of the vibron,  revealing a set of narrow additional sidebands which can increase the absorption of the molecule by orders of magnitude. The narrow sidebands stem from transitions on the zero-phonon line in combination with the simultaneous absorption of infrared photons (see Fig.~\ref{fig3}(c) for possible processes contributing to first vibronic sideband). Dephasing of the zero-phonon line transition (see Fig.~\ref{fig3}(b)) increases the width and reduces the height of these driving-induced sidebands.  For $\tilde{\gamma}\ll\Gamma$, the intensity of the zero-phonon line ($S(\Delta_p=0)$) can be approximated by ($n=m=\ell=0$, assuming all other combinations $n\nu+(\ell+m)\omega_d=0$ to be much smaller)
\begin{align}
I_{\text{ZPL}}=\frac{\eta_p^2}{\gamma\tilde{\gamma}}e^{-\lambda^2}\mathcal{J}_0\!\left(\frac{2\lambda\eta_d}{\omega_d}\right)^2\BesselJ_0\left(2\lambda|\beta|\right)^2.
\end{align}
For small arguments of the Bessel functions, the effective Franck-Condon factor describing the oscillator strength of the zero-phonon line transition can be approximated by $\text{exp}\!\left[{-\lambda^2(1+2|\beta|^2+2\eta_d^2/\omega_d^2)}\right]$. Assuming a narrow zero-phonon linewidth $ \tilde{\gamma}\ll\Gamma$, the main contribution of the infrared drive in Eq.~(\ref{modabsorption}) stems from $n=0$ and the intensity of the driving-induced sideband $I_d$ at e.g.~$\Delta_p=\nu$ (normalized by the intensity of the bare vibronic sideband $I_b$) can be approximated by ${I_d}/{I_b}\approx\eta_d^2/(4\tilde{\gamma}\Gamma)$ ($\BesselJ_1(x)^2\approx (x/2)^2$ for $x\ll 1$), i.e., it increases with the driving amplitude $\eta_d$ and is mitigated by dephasing $\gamma_{\phi}$ and vibrational relaxation $\Gamma$.\\
\indent An interesting special case is given if one considers off-resonant driving of the vibration (e.g.,~$\omega_d <\nu$) and consequently $|\beta|^2\approx 0$. The vibrational mode of the molecule then essentially undergoes free evolution: $\dot{p}=-2\Gamma p-\nu q +\xi$, and the displacement correlation function becomes stationary and reads
\begin{align}
\braket{\mathcal{D}(t)\mathcal{D}^\dagger (t')}=e^{-\lambda^2}e^{\lambda^2\mathrm{exp}\left[-(\Gamma+i\nu)(t-t')\right]}.
\end{align}
In this case, the only effect of the drive is the modulation of the electronic transition frequency  $\tilde{\omega}_0=\omega_0+2\lambda\eta_d\cos(\omega_d t)$ and the absorption spectrum simplifies to
\begin{align}
S(\Delta_p)&=\frac{\eta_p^2}{\gamma}\sum_{\substack{n=0\\ m=-\infty}}^\infty\!\frac{s_n^\lambda(\tilde{\gamma}\!+\!n\Gamma)\mathcal{J}_m\!\left(\frac{2\lambda\eta_d}{\omega_d}\right)^{\!2}}{(\tilde{\gamma}+n\Gamma)^2+(\Delta_p\!-\!n\nu-\!m\omega_d)^2}.
\end{align}
The intensities of the driving-induced sidebands are now only determined by the  vibronic coupling strength $\lambda$ as well as by the ratio between driving amplitude and driving frequency $\eta_d/\omega_d$. Similar terms appear for frequency-modulated pure two-level systems such as atoms or superconducting qubits \cite{silveri2017quantum, wilson2007coherence}.  Interestingly, the Bessel functions become zero at some values (e.g., $\BesselJ_0(z)^2\approx 0$ for $z\approx 2.41$, see Fig.~\ref{fig2}(c)) which can be used to decouple specific sidebands completely from the driving field. The corresponding Floquet states $\ket{e;m}$ are then dark, a phenomenon also known as coherent destruction of tunneling (CDT) \cite{grossmann1991coherent}.\\

\subsection{Modification of vibronic coherence}

\begin{figure}[t]
\includegraphics[width=1.01\columnwidth]{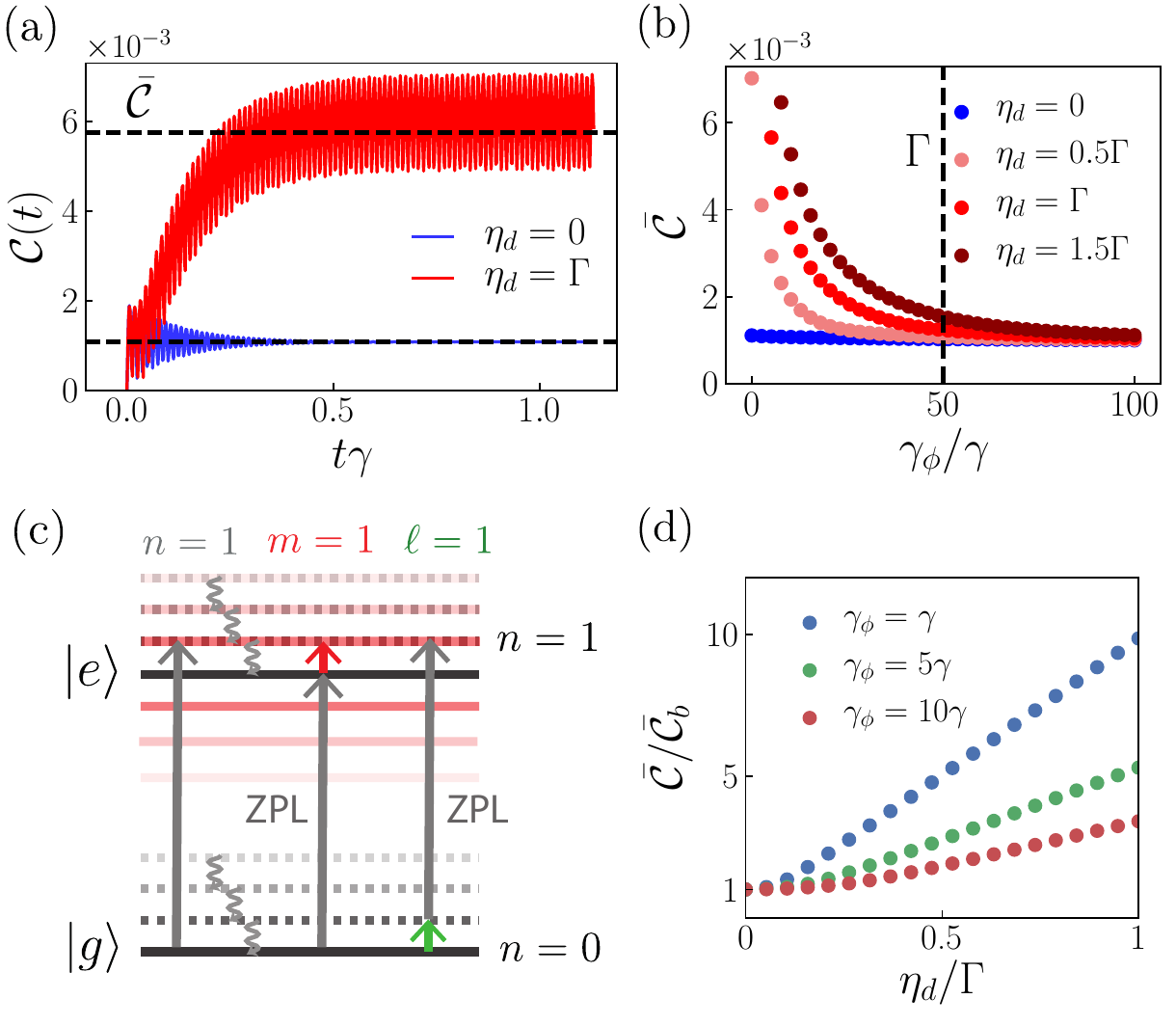}
\caption{\emph{Sideband coherence}. (a) Coherence $\mathcal{C}(t)$ of the first vibronic sideband at $\Delta_p=\nu$ with and without (resonant) driving of the vibrational mode. The black dashed lines show the time-averaged coherences $\bar{\mathcal{C}}$. Parameters: $\gamma_\phi=2\gamma$, $\Gamma=50\gamma$, $\lambda=0.2$, $\eta_p=0.1\gamma$. (b) Effect of dephasing on $\bar{\mathcal{C}}$ at the first vibronic sideband $\Delta_p=\nu$  for different driving strengths for the same parameters as in (a), except that now $\gamma_\phi$ is varied.  The dashed vertical line shows the magnitude of $\Gamma$. (c) Processes contributing to the generation of coherence in the first vibronic sideband: direct transition (grey), transition via electronic Floquet state (red) and transition due to absorption of a vibrational quantum (green). (d) Total coherence $\bar{\mathcal{C}}$ of the first sideband normalized to the bare vibronic coherence $\bar{\mathcal{C}}_b$ as a function of the driving strength $\eta_d$ for different strengths of dephasing.}
\label{fig4}
\end{figure}

In principle, due to their strong internal vibronic coupling (the dimensionless coupling $\lambda$ can be on the order of unity), molecules could be utilized as ideal quantum opto-mechanical systems already in the strong coupling regime, where light can be mapped onto motion (via the electronic degree of freedom) and vice versa~\cite{roelli2016molecular, neuman2019quantum, benz2016single, roelli2020molecular}. This is however hampered by the quick vibrational relaxation process $\Gamma$ that reduces the vibrational coherence and leads to low mechanical $Q$-factors $Q=\nu/\Gamma$ for the opto-vibrational interaction. The detrimental effect can be mitigated by the continuous driving of vibrations leading to narrower transitions more robust with respect to losses (as illustrated in Fig.~\ref{fig3}(a,b)). One can quantify this increased robustness by defining a measure of coherence for the electronic part in terms of the standard Pauli operators as
\begin{align}
\mathcal{C}(t)=\sqrt{\braket{\sigma_x}^2+\braket{\sigma_y}^2},
\end{align}
which can be expressed in terms of raising and lowering operators as $\mathcal{C}(t)=|\braket{\sigma(t)}|+|\braket{\sigma^\dagger (t)}|=2|\braket{\sigma (t)}|$ \cite{baumgratz2014quantifying}. For the driven system, we can define the time-averaged coherence at times $t\gg1/\tilde{\gamma}$
\begin{align}
\bar{\mathcal{C}}=\lim_{T \to \infty}\frac{1}{T}\int_t^{t+T} dt' \,\mathcal{C}(t'),
\end{align}
which is illustrated in Fig.~\ref{fig4}(a). In Fig.~\ref{fig4}(b) we plot the time-averaged coherence at the first vibronic sideband ($\Delta_p=\nu$) for different driving strengths as a function of dephasing $\gamma_\phi$. We can see that driving can greatly enhance the coherence of the sidebands for $\tilde{\gamma}\ll\Gamma$, i.e., if the zero-phonon linewidth is narrow compared to the linewidth of the vibronic sidebands $\tilde{\gamma}+n\Gamma$ (we focus on the first vibronic sideband $n=1$). If $\tilde{\gamma}$ becomes on the order of $\Gamma$, the sideband-coherence of the driven system gets close to that of the undriven system. From the time evolution of $\braket{\sigma(t)}$, one can estimate the driving-induced contributions (additional to $n=1$) to the coherence of the first vibronic sideband for $\omega_d=\nu$ at large times $t\gg1/\tilde{\gamma}$ as
\begin{align}
\braket{\sigma(t)}_d&=\frac{\eta_p}{\tilde{\gamma}} e^{-\lambda^2}e^{i\nu t}\\\nonumber
&\times\Big[\BesselJ_1\!\left(\frac{2\lambda\eta_d}{\omega_d}\right)\!\BesselJ_0\!\left(2\lambda|\beta|\right)\!+\!\BesselJ_0\!\left(\frac{2\lambda\eta_d}{\omega_d}\right)\!\BesselJ_1\!\left(2\lambda|\beta|\right)\Big],
\end{align}
which has to be compared with the bare vibronic contribution of
\begin{align}
\braket{\sigma}_b=\eta_p e^{-\lambda^2}\frac{\lambda^2}{\tilde{\gamma}+\Gamma}\BesselJ_0\left(\frac{2\lambda\eta_d}{\omega_d}\right)\BesselJ_0\left(2\lambda|\beta|\right).
\end{align}
The respective processes are illustrated in Fig.~\ref{fig4}(c), showing either a direct transition (with $\tilde{\gamma}+\Gamma$) or a transition on the zero-phonon line (with $\tilde{\gamma}$) in combination with the absorption of an infrared photon. In Figure \ref{fig4}(d) we plot the coherence $\bar{\mathcal{C}}/\bar{\mathcal{C}}_b$ (normalized by the coherence of the undriven system $\bar{\mathcal{C}}_b$) of the first vibronic sideband as a function of the driving strength, revealing a largely linear dependence between the coherence and the driving strength ($\BesselJ_1(x)\approx x/2$ for small $x$).

\subsection{\label{sec:level3}Purcell-induced suppression of sidebands}
We now turn to the situation depicted in Fig.~\ref{fig1} and consider a molecule coupled to a confined quantized electromagnetic in the infrared. The molecule is only indirectly coupled to a laser drive that feeds the cavity field with an amplitude $\eta_d^c=\sqrt{{2\mathcal{P}\kappa}/{(\hbar\omega_d)}}$ ($\mathcal{P}$ is the laser power) and frequency $\omega_d$.  The coupling to the quantized cavity mode \red{with photon annihilation and creation operators $a$ and $a^\dagger$} is described by the Hamiltonian
\begin{align}
\mathcal{H}_{\text{cav}}=g(a^\dagger+a)(b^\dagger+b).
\end{align}
The coupling is given by $g=\boldsymbol{\epsilon}_c\!\cdot\!\boldsymbol{\epsilon}_m\mathcal{E}_0 (\partial\braket{{\mu}}/\partial q)_0 q_{\text{zpm}}$ with the zero point field amplitude $\mathcal{E}_0=\sqrt{\hbar\omega_c/(2\varepsilon_0 \mathcal{V})}$ ($\mathcal{V}$ is the mode volume, $\varepsilon_0$ is the vacuum permittivity). Here, the unit vectors $\boldsymbol{\epsilon}_c$ and $\boldsymbol{\epsilon}_m$ account for the polarizations of electric field and molecular dipole, respectively. The dynamics of the system can be followed at the level of quantum Langevin equations for mechanical and optical degrees of freedom as well as for the polaron operator $\tilde{\sigma}$ which read
\begin{subequations}
\begin{align}
\dot{q}&=\nu p,\\
\dot{p}&=-2\Gamma p -\nu q-\sqrt{2}g(a^\dagger+a)+\xi(t),\\
\dot{a}&=-(\kappa+i\omega_c)a-i\sqrt{2}gq+\sqrt{2\kappa} a_{\text{in}}\!+\!\eta_d^c\cos(\omega_d t),\\
\dot{\tilde{\sigma}}&=-(\tilde{\gamma}-i\Delta_p)\tilde{\sigma}-2i\lambda g(a^\dagger+a)\tilde{\sigma}+\tilde{\Sigma}_{\text{in}},
\end{align}
\end{subequations}
where the coupling to the cavity field again leads to a shift of the polaronic transition frequency.\\
\indent Assuming resonant driving $\omega_c=\omega_d$, we can identify a first advantage of using the cavity as an increase in the driving strength brought on by the cavity finesse. This can be seen by an elimination of the cavity field to find an effective equation for the polaron operator
\begin{figure}[b]
\includegraphics[width=0.85\columnwidth]{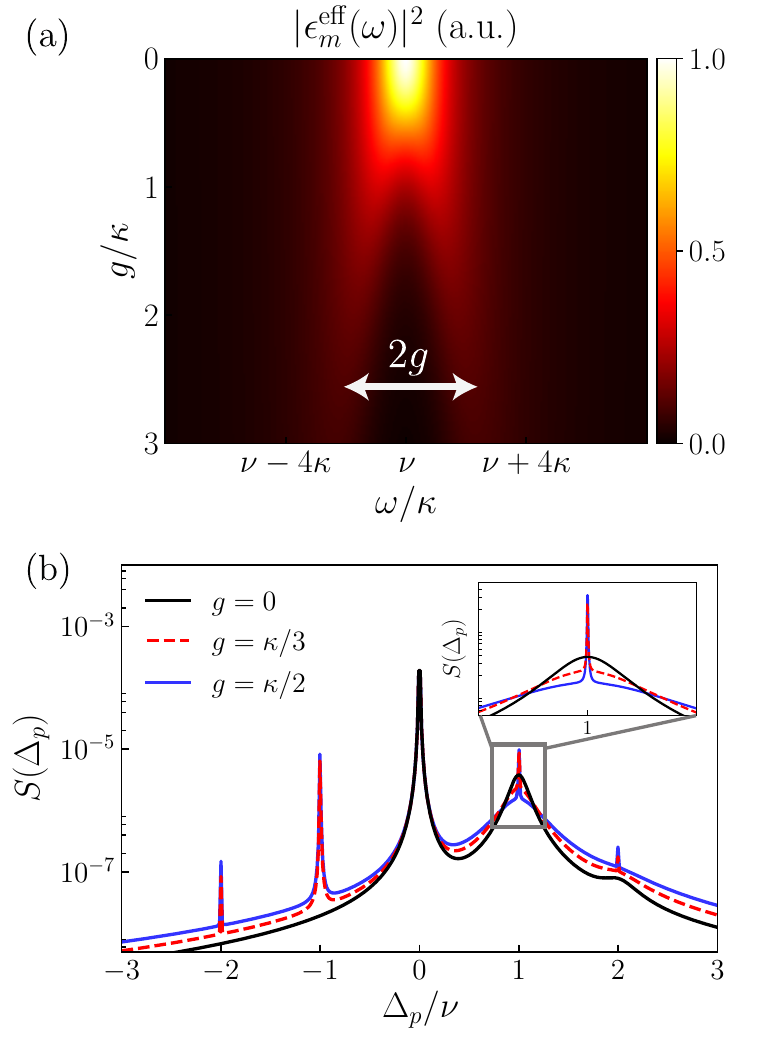}
\caption{\emph{Cavity-modified absorption}. (a)  Effective mechanical susceptibility as a function of $g$ and $\omega$ for $\Gamma=\kappa$. For $g>\kappa$, the susceptibility shows a mode splitting of approximately $2g$. (b) Cavity-modified optical absorption spectrum for different values of $g$ in the weak coupling regime. Parameters: $\lambda=0.2$, $\Gamma=100\gamma$, $\kappa=600\gamma=6\Gamma$, $\gamma_\phi=0$, $\eta_d^c=0.8\kappa$, $\eta_p=0.1\gamma$.}
\label{fig5}
\end{figure}
\begin{align}
\label{resonantsigma}
\dot{\tilde{\sigma}}&=-(\tilde{\gamma}-i\Delta_p)\tilde{\sigma}-2i\lambda g\frac{\eta_d^c}{\kappa}\cos(\omega_d t)\tilde{\sigma}+\tilde{\Sigma}_{\text{in}},
\end{align}
similar to Eq.~\eqref{eqnsfreespace} but with a modified drive strength. Similarly to the case of direct driving, the periodic modulation of the electronic transition frequency $\tilde{\omega}_0=\omega_0+2\lambda g\eta_d^c\cos(\omega_d t)/\kappa$ leads to sidebands proportional to $\BesselJ_m\left(\frac{2\lambda g\eta_d^c}{\kappa\omega_d}\right)^2$. However, at a difference to the direct driving case, the free space coupling $\eta_d$ has now been replaced by a cavity enhanced strength $g\eta_d^c /\kappa$. To estimate the gain in this approach, let us assume a free space propagating infrared laser beam with transverse area $\mathcal{S}$ and amplitude $\mathcal{E}$ such that $\mathcal{P}=\epsilon_0 c \mathcal{E}^2 \mathcal{S}/2$. If the molecule were to be under direct laser illumination, the free space Rabi frequency could be written as $\eta_d=\mu \mathcal{E}/\hbar$. In the cavity, using $\kappa=\pi c/(L\mathcal{F})$ ($\mathcal{F}$ is the cavity finesse and $L$ the cavity length) and estimating $g=\mu/\hbar \sqrt{\hbar \omega_c/(2\epsilon_0 L \mathcal{S}})$ we can then estimate the ratio $(g\eta_d^c/\kappa)/\eta_d=\sqrt{\mathcal{F}/(2\pi)}$. A first advantage of using a cavity is therefore insured when employing a high-finesse resonator, providing an enhancement proportional to $\sqrt{\mathcal{F}}$ of all previously discussed effects.\\
\noindent
\indent The second advantage presented by the cavity is an effective Purcell increase of the vibrational relaxation rate when large cooperativities $ \mathcal{C}_\text{cav}=g^2/(\kappa \Gamma)>1$ are employed. To illustrate this effect, we follow a standard optomechanical approach \cite{aspelmeyer2014cavity, dantan2008self}, where we incorporate the effect of the infrared field onto the vibrations as an additional cooling bath. This can be seen in the frequency space as an optical force $p(\omega)=\epsilon_m (\omega)[\xi(\omega)+F_{\text{IR}}(\omega)]$ (with  $F_{\text{IR}}(\omega)=-\sqrt{2}g[a^\dagger(\omega)+a(\omega)]$) added to the intrinsic Brownian noise term $\xi(\omega)$. The bare susceptibility term $\epsilon_m(\omega)=i\omega/(\omega^2+2i\Gamma\omega-\nu^2)$ describes the response of the vibron to the surrounding bath of oscillators.
The additional effect of the infrared optical bath is to give rise to a modified mechanical susceptibility
\begin{align}
(\epsilon_m^{\text{eff}})^{-1}(\omega)=[\epsilon_m^{-1}+2g^2\frac{\nu}{\omega}\left(\epsilon_c(\omega)-\epsilon_c^*(-\omega)\right)],
\end{align}
by introducing the effect of the cavity via its susceptibiity $\epsilon_c(\omega)=1/[{i(\omega_c-\omega)+\kappa}]$. This is the standard picture of sideband cooling where an effective damping rate can be computed (evaluated at $\omega=\nu$)
\begin{align}
\Gamma_\text{IR}=\frac{g^2\kappa}{\kappa^2+(\omega_d-\nu)^2}-\frac{g^2\kappa}{\kappa^2+(\omega_d+\nu)^2}.
\end{align}
The resolved sideband condition requiring $\omega_d=\nu$ and $\kappa\ll\nu$ gives rise to a simple expression for the optically induced cooling rate $\Gamma_\text{IR}=g^2/\kappa$. Adding this expression to the bare vibrational damping rate we obtain $\Gamma_\text{eff}=\Gamma+\Gamma_\text{IR}=\Gamma(1+\mathcal{C}_\text{cav})$. This reveals this effect as stemming from a Purcell modification of the damping rate owing to coupling to a lossy cavity.\\
\indent Going towards the strong coupling regime, Fig.~\ref{fig5}(a) illustrates the effective mechanical susceptibility as a function of $g$, revealing a mode splitting at the onset of strong coupling $g>\kappa$. To calculate the cavity-modified absorption spectrum, one then has to again evaluate the momentum correlation function $\braket{p(t)p(t')}$ which can be obtained similarly to the free space case by calculating the correlations in frequency domain and then transforming back to the time domain. The stationary part of the momentum correlation function (full expression specified in Appendix) in the weak coupling regime is given by
 \begin{align}
 \braket{p(t)p(t')}_{\text{st}}=\frac{e^{-(\tilde{\Gamma}+i\nu)(t-t')}\!}{2}+|\beta_{\text{c}}|^2\cos(\omega_d (t-t')),
 \end{align}
with the cavity-induced vibrational occupation  $|\beta_{\text{c}}|^2=g^2(\eta_d^c)^2|\epsilon_m^{\text{eff}}(\omega_d)|^2|\epsilon_c (\omega_d)^2|$.
After integrating Eq.~(\ref{resonantsigma}) and calculating $\braket{\sigma (t)}$, the absorption spectrum (assuming weak coupling $g<\kappa$) for $\gamma\ll\omega_d$ can be expressed as
\begin{align}
S(\Delta_p)&\!=\\\nonumber
&\!\frac{\eta_p^2}{\gamma}\!\sum_{\substack{n=0\\ m,\ell=-\infty}}^\infty\!\frac{s_n^\lambda (\tilde{\gamma}\!+\!n\tilde{\Gamma})\mathcal{J}_m(z)^2\BesselJ_\ell\left(2\lambda|\beta_{\text{c}}|\right)^2}{(\tilde{\gamma}\!+\!n\tilde{\Gamma})^2+(\Delta_p\!-\!n\nu-\!(m\!+\!\ell)\omega_d)^2}.
\end{align}
with $z=2\lambda g\eta_d^c/(\kappa\omega_d)$.  In the case of complete resonance $\omega_d=\omega_c=\nu$, one can approximate $|\beta_{\text{c}}|^2\approx  [g\eta_d^c/(2\kappa\Gamma)]^2$. The previously discussed Purcell effect is clearly illustrated in the optical spectrum in  Fig.~\ref{fig5}(b) which shows a a leveling of the floor around the driven vibrational sideband at $\Delta_p=\nu$. For increasing values of the cooperativity the amplitude of the sidebands is increasing (owing to the increase in $g$) while at the same time, absorption outside this narrow interval is strongly suppressed.

\section{\label{sec:level4}Conclusions and outlook}
The optical properties of infrared active molecules can be strongly modified by direct coherent drive of vibrational transitions. We have introduced an analytical treatment to these processes based on an open system approach using quantum Langevin equations. Two main results of the paper are that both amplitudes and linewidths of particular vibrational sidebands can be externally tuned by tuning the strength and frequency of the infrared drive. These effects can find applications in improving the vibronic coherence, i.e. creating vibronic transitions robust with respect to vibrational relaxation. Moreover, the additional drive through an infrared cavity resonant to vibrational transitions can benefit from the cavity-induced Purcell effect and lead to the suppression of undesired broad vibronic sidebands. While the current investigations are restricted to the level of one molecule, future endeavors will see the extension of these calculations to the level of collective coupling of a mesoscopic molecular ensemble to an infrared cavity. \red{We will also aim at extending our model to the anharmonic case and to the case of avoided crossings where the electron-vibron coupling would see position dependence}. Our formalism could also be applied to the near field energy transfer between two molecules (the so-called F{\"o}rster resonance energy transfer) where one could make use of the sideband control to increase the spectral overlap and therefore also the energy transfer rate between molecules. Similar ideas have already been discussed for the modulation of purely electronic degrees of freedom \cite{kohler2005driven, thanphuc2018control}.

\section*{Supplementary Material}
See Supplementary Material for more detailed derivations of the results in the main text.

\begin{acknowledgments}
We acknowledge financial support from the Max Planck Society and from the German Federal Ministry of Education and Research, co-funded by the European Commission (project RouTe), project number 13N14839 within the research program "Photonik Forschung Deutschland". We would like to acknowledge fruitful discussions with Christian Sommer and we thank Rinat Tyumenev for useful comments on the manuscript.
\end{acknowledgments}

\section*{Data availability}

The data that support the findings of this study are available from the corresponding author
upon reasonable request.

\bibliography{REFinfrared}

\pagebreak
\onecolumngrid
\appendix

\preprint{AIP/123-QED}

\maketitle
\onecolumngrid
\section{Free space driving: Full equations of motion}
\label{freespacedriving}

For a single molecular vibrational mode periodically driven by $\mathcal{H}_d=\eta_d\cos(\omega_d t)(b^\dagger+b)$, the full equations of motion for mechanical and electronic degrees of freedom are given by
\begin{subequations}
\begin{align}
\dot{q}&=\nu p,\\
\dot{p}&=-2\Gamma p-\nu q-\sqrt{2}\eta_d\cos(\omega_d t)-\sqrt{2}\lambda\nu\sigma^\dagger\sigma+\xi(t),\\\nonumber
\dot{\sigma}&=-(\tilde{\gamma}-i\Delta_p+i\lambda^2\nu)\sigma+i\sqrt{2}\lambda q\sigma+\eta_p(1-2\sigma^\dagger\sigma)+\sqrt{2\gamma}\sigma_{\text{in}}(1-2\sigma^\dagger\sigma)+\sqrt{2\gamma_\phi}\left((\sigma^\dagger\sigma)_{\text{in}}\sigma-\sigma(\sigma^\dagger\sigma)_{\text{in}}\right),
\end{align}
\end{subequations}
with $\tilde{\gamma}=\gamma+\gamma_\phi$. Using that the equation of motion for the displacement operator $\mathcal{D}^\dagger=e^{i\lambda\sqrt{2} p}$ is given by \cite{suzuki1997quantum}
\begin{align}
\dot{\mathcal{D}}^\dagger&=i\sqrt{2}\lambda\mathcal{D}^\dagger\int_{0}^1 ds e^{-is\sqrt{2}\lambda p}\dot{p}(t)e^{is\sqrt{2}\lambda p}\\\nonumber
&=-2\sqrt{2}i\Gamma\lambda\mathcal{D}^\dagger p-\lambda i\nu\sqrt{2}\mathcal{D}^\dagger q +i\nu\lambda^2\mathcal{D}^\dagger+2i\nu\lambda^2\mathcal{D}^\dagger\sigma^\dagger\sigma-2i\lambda\eta_d\cos(\omega_d t)\mathcal{D}^\dagger+\sqrt{2}i\lambda\mathcal{D}^\dagger\xi,
\end{align}
under the assumptions of weak probe fields ($\eta_p\ll\gamma$ and therefore $\braket{\sigma^\dagger\sigma}\ll 1$), the equations of motion for mechanical degrees of freedom and polaron operator $\dot{\tilde{\sigma}}=\dot{\sigma}\mathcal{D}^\dagger+\sigma\dot{\mathcal{D}}^\dagger$ can be simplified to \cite{reitz2019langevin}
\begin{subequations}
\begin{align}
\dot{q}&=\nu p,\\
\dot{p}&=-2\Gamma p-\nu q-\sqrt{2}\eta_d\cos(\omega_d t)+\xi(t),\\
\dot{\tilde{\sigma}}&=-(\tilde{\gamma}-i\Delta_p)\tilde{\sigma}-2i\lambda\eta_d\cos(\omega_d t)\tilde{\sigma}+\eta_p\mathcal{D}^\dagger+\sqrt{2\gamma}\tilde{\sigma}_{\text{in}}+\sqrt{2\gamma_\phi}\left((\sigma^\dagger\sigma)_{\text{in}}\sigma-\sigma(\sigma^\dagger\sigma)_{\text{in}}\right)]\mathcal{D}^\dagger,
\label{polaronfreespace}
\end{align}
\end{subequations}
where we  made use of the commutation relations $[p,\mathcal{D}^\dagger]=0$, $[q,\mathcal{D}^\dagger]=-\sqrt{2}\lambda\mathcal{D}^\dagger$. Additionally, the population of the electronic state evolves according to
\begin{align}
\frac{d (\sdag \s)}{dt} &=-2\gamma\sigma^\dagger\sigma+2\eta_p\Re\{\sigma (t)\}+\sqrt{2\gamma}\left[\sdag \siginp+\siginpd \s\right].
\end{align}
The formal solution of the time evolution of the dipole operator (initial condition at $-\infty$) is given by:
\begin{align}
\label{sigmaevolution}
\braket{\sigma(t)}&=\eta_p\int_{-\infty}^t dt' e^{-(\tilde{\gamma}-i\Delta_p)(t-t')}e^{-2i\lambda\eta_d\int_{t'}^t ds\cos(\omega_d s)}\braket{\mathcal{D}(t)\mathcal{D}^\dagger (t')}=\\\nonumber
&=\eta_p\int_{-\infty}^t dt' e^{-(\tilde{\gamma}-i\Delta_p)(t-t')}e^{-2i\lambda\eta_d\left[\sin(\omega_d t)-\sin(\omega_d t')\right]/\omega_d}\braket{\mathcal{D}(t)\mathcal{D}^\dagger (t')}.
\end{align}

\subsection{Off-resonant driving}
\label{offresonantdriving}

Let us first consider a very off-resonant drive of the vibrations (e.g.,~$\omega_d\ll\nu$) such that the vibrational mode itself is not excited and remains in the ground state. The only effect is then the modulation of the electronic transition frequency and one can then assume free evolution of the vibrations $\dot{p}=-2\Gamma p-\nu q+\xi(t)$. The displacement correlation function for times $t\geq t'$ is then given by (at zero temperature)
\begin{align}
\label{displacementcorr}
\braket{\mathcal{D}(t)\mathcal{D}^\dagger (t')}&=\braket{e^{-i\sqrt{2}\lambda p(t)}e^{i\sqrt{2}\lambda p(t')}}=\braket{e^{-i\sqrt{2}\lambda(p(t)-p(t'))}}e^{\lambda^2[p(t),p(t')]}\\\nonumber
&=\braket{\sum_m \frac{(-i\sqrt{2}\lambda)^{2m}}{(2m)!}\left(p(t)-p(t')\right)^{2m}}e^{\lambda^2[p(t),p(t')]}=e^{-\lambda^2\braket{(p(t)-p(t'))^2}}e^{\lambda^2\braket{[p(t),p(t')]}}=\\\nonumber
&=e^{-\lambda^2\left(\braket{p(t)^2}+\braket{p(t')^2}-2\braket{p(t)p(t')}\right)}=e^{-2\lambda^2 (\braket{p^2}-\braket{p(t)p(t')})}=e^{-\lambda^2}e^{\lambda^2 e^{-(\Gamma+i\nu)(t-t')}},
\end{align}
where we made use of the Isserlis' theorem [$(2m)!/(2^m m!)$ partitions in the sum] and used that the momentum variance is stationary $\braket{p^2}=\braket{p(t')^2}=\braket{p(t)^2}=1/2$. With this, one can calculate the integral in Eq.~(\ref{sigmaevolution}) by means of the Jacobi-Anger expansion $e^{iz\sin\Phi}=\sum_{m=-\infty}^\infty \mathcal{J}_m (z)e^{i m\Phi}$ ($\BesselJ_m (z)$ are the Bessel functions of the first kind)
\begin{align}
\braket{\sigma(t)}&=\eta_p e^{-\lambda^2} e^{-2i\lambda\eta_d\sin(\omega_d t)/\omega_d}\int_{-\infty}^t dt' e^{-(\tilde{\gamma}-i\Delta_p)(t-t')}\sum_{n=0}^\infty\frac{\lambda^{2n}}{n!}e^{-n(\Gamma+i\nu)(t-t')}\sum_{m=-\infty}^\infty \mathcal{J}_m\left(\frac{2\lambda\eta_d}{\omega_d}\right)e^{i m \omega_d t'}=\\\nonumber
&=\eta_p e^{-\lambda^2}\sum_{n=0}^\infty\sum_{m=-\infty}^\infty\frac{\lambda^{2n}}{n!}\frac{\mathcal{J}_m\left(\frac{2\lambda\eta_d}{\omega_d}\right)e^{-i\eta_d\sin(\omega_d t)/\omega_d}e^{im\omega_d t}}{(\tilde{\gamma}+n\Gamma)-i(\Delta_p-n\nu-m\omega_d)}.
\end{align}
While the dipole moment is non-stationary, the excited-state population becomes stationary and one finds
\begin{align}
\braket{\sigma^\dagger\sigma (t)}&=2\eta_p\Re\left[\int_{-\infty}^t dt' e^{-2\gamma(t-t')}\braket{\sigma(t')}\right]=\\\nonumber
&=2\eta_p^2 \Re\left[\sum_{n=0}^\infty\sum_{m=-\infty}^\infty\sum_{\ell=-\infty}^\infty \frac{\lambda^{2n}}{n!}\frac{e^{-i(\ell-m)\omega_d t}}{2\gamma-i(\ell-m)\omega_d}\frac{\mathcal{J}_m\left(\frac{2\lambda\eta_d}{\omega_d}\right)\mathcal{J}_\ell\left(\frac{2\lambda\eta_d}{\omega_d}\right)}{(\tilde{\gamma}+n\Gamma)-i(\Delta_p-n\nu-m\omega_d)}\right]=\\\nonumber
&=\frac{\eta_p^2}{\gamma}\sum_{n=0}^\infty\sum_{m=-\infty}^\infty\frac{\lambda^{2n}}{n!}\frac{(\tilde{\gamma}+n\Gamma)\mathcal{J}_m\left(\frac{2\lambda\eta_d}{\omega_d}\right)^2}{(\tilde{\gamma}+n\Gamma)^2+(\Delta_p-n\nu-m\omega_d)^2},
\end{align}
where for $\gamma\ll\omega_d$ one can approximate $\sum_{\ell=-\infty}^\infty \frac{e^{-i(\ell-m)\omega_d t}}{2\gamma-i(\ell-m)\omega_d}f_{\ell}\approx\sum_{\ell=-\infty}^\infty1/(2\gamma)\delta_{m\ell}f_{\ell}=1/(2\gamma)f_m$.

\subsection{Resonant driving}

In the case of resonant driving $\omega_d \approx\nu$, the vibrations will be driven into a coherent state with average vibron occupancy given by $\braket{b^\dagger b}=|\beta|^2=\frac{(\eta_d/2)^2}{\Gamma^2+(\omega_d-\nu)^2}$ and $\braket{p(t)}\neq 0$. The displacement correlation function then becomes non-stationary and is given by  (again applying the Isserlis' theorem)
\begin{align}
\braket{\mathcal{D}(t)\mathcal{D}^\dagger (t')}&=\braket{e^{-i\sqrt{2}\lambda p(t)}e^{i\sqrt{2}\lambda p(t')}}=\braket{e^{-i\sqrt{2}\lambda(p(t)-p(t'))}}e^{\lambda^2[p(t),p(t')]}\\\nonumber
&=\braket{\sum_m \frac{(-i\sqrt{2}\lambda)^{2m}}{(2m)!}\left(P(t,t')-\braket{P(t,t')}\right)^{2m}}e^{\lambda^2[p(t),p(t')]}e^{-i\sqrt{2}\lambda\braket{P(t,t')}}\\\nonumber
&=e^{-\lambda^2\braket{(P(t,t')-\braket{P(t,t')})^2}}e^{\lambda^2\braket{[p(t),p(t')]}}e^{-i\sqrt{2}\lambda\braket{P(t,t')}}=e^{-\lambda^2}e^{\lambda^2 e^{-(\Gamma+i\nu) (t-t')}}e^{-i\sqrt{2}\lambda\braket{P(t,t')}}.
\end{align}
where we denoted $P(t,t')=p(t)-p(t')$, made use of the fact that the variance of coherent states is identical to vacuum states $\braket{p(t)^2}-\braket{p(t)}^2=\frac{1}{2}$ and used furthermore that the momentum correlation function is given by (for derivation see Appendix \ref{momcorr} below)
\begin{align}
\braket{p(t)p(t')}&=\frac{1}{2}\left(\beta^*\beta^* e^{i\omega_d (t+t')}+\beta\beta e^{-i\omega_d(t+t')}+2|\beta|^2\cos\left(\omega_d(t-t')\right)+e^{-(\Gamma+i\nu)(t-t')}\right),
\end{align}
with $\beta=(\eta_d/2)/\left[\Gamma-i(\omega_d-\nu)\right]$. Using that $\braket{p(t)}=\left(\beta^* e^{i\omega_d t}+\beta e^{-i\omega_d t}\right)/\sqrt{2}$, on resonance this becomes ($\beta\approx\beta^*$)
\begin{align}
\braket{\mathcal{D}(t)\mathcal{D}^\dagger (t')}=e^{-\lambda^2}e^{\lambda^2 e^{-(\Gamma+i\nu) (t-t')}}e^{-2i\lambda|\beta|\left(\cos(\omega_d t)-\cos(\omega_d t')\right)}.
\end{align}
The dipole moment can then be expressed as
\begin{align}
\braket{\sigma(t)}&=\eta_p\int_{-\infty}^t dt' e^{-(\tilde{\gamma}-i\Delta_p)(t-t')}e^{-2i\lambda\eta_d\left[\sin(\omega_d t)-\sin(\omega_d t')\right]/\omega_d}e^{-\lambda^2}e^{\lambda^2 e^{-(\Gamma+i\nu) (t-t')}}e^{-2i\lambda|\beta|\left(\cos(\omega_d t)-\cos(\omega_d t')\right)}\\\nonumber
&=\eta_p e^{-\lambda^2} \sum_{n=0}^\infty\sum_{m,\ell=-\infty}^\infty \frac{\lambda^{2n}}{n!}\frac{\BesselJ_m\left(\frac{2\lambda\eta_d}{\omega_d}\right)i^\ell\BesselJ_\ell\left(2\lambda|\beta|\right)e^{i(m+\ell)\omega_d t}}{(\tilde{\gamma}+n\Gamma)-i(\Delta-n\nu-m\omega_d-\ell\omega_d)}e^{-2i\lambda\eta_d\sin(\omega_d t)/\omega_d}e^{-2i\lambda|\beta|\cos(\omega_d t)}.
\end{align}
The steady-state population $\braket{\sigma^\dagger\sigma}$ in the limit $\gamma\ll\omega_d$ can then be calculated as
\begin{align}
\braket{\sigma^\dagger\sigma (t)}&=\\\nonumber
&2{\eta_p^2}e^{-\lambda^2}\sum_{n=0}^\infty \sum_{m,\ell,h,k=0}^\infty\frac{\lambda^{2n}}{n!} \frac{e^{i(m+\ell+h-k)\omega_d t}}{2\gamma-i(k-m-\ell-h)\omega_d}\Re\left\{\frac{i^\ell i^h \BesselJ_m\left(\frac{2\lambda\eta_d}{\omega_d}\right)\BesselJ_k\left(\frac{2\lambda\eta_d}{\omega_d}\right)\BesselJ_\ell\left(2\lambda|\beta|\right)\BesselJ_h\left(-2\lambda|\beta|\right)}{(\tilde{\gamma}+n\Gamma)-i(\Delta-n\nu-m\omega_d-\ell\omega_d)}\right\}\\\nonumber
&\approx \frac{\eta_p^2}{\gamma}e^{-\lambda^2}\sum_{n=0}^\infty \sum_{m,\ell,k=-\infty}^\infty\frac{\lambda^{2n}}{n!} \Re\left\{\frac{i^{k-m} \BesselJ_m\left(\frac{2\lambda\eta_d}{\omega_d}\right)\BesselJ_k\left(\frac{2\lambda\eta_d}{\omega_d}\right)\BesselJ_\ell\left(2\lambda|\beta|\right)\BesselJ_{k-m-\ell}\left(-2\lambda|\beta|\right)}{(\tilde{\gamma}+n\Gamma)-i(\Delta-n\nu-m\omega_d-\ell\omega_d)}\right\}\\\nonumber
&=\frac{\eta_p^2}{\gamma}e^{-\lambda^2}\sum_{n=0}^\infty\sum_{m,\ell=-\infty}^\infty\frac{\lambda^{2n}}{n!}\frac{(\tilde{\gamma}+n\Gamma)\mathcal{J}_m\left(\frac{2\lambda\eta_d}{\omega_d}\right)^2 \BesselJ_\ell\left(2\lambda|\beta|\right)^2}{(\tilde{\gamma}+n\Gamma)^2+(\Delta_p-n\nu-(m+\ell)\omega_d)^2}.
\end{align}

\section{Momentum correlation function}
\label{momcorr}

For the calculation of the momentum correlation function it is convenient to go to Fourier space (using the definition $O(\omega)=\mathcal{F}\left\{O(t)\right\}=\frac{1}{\sqrt{2\pi}}\int_{-\infty}^\infty d t O(t) e^{i\omega t}$):
\begin{align}
p(\omega)=\epsilon_m(\omega)\left[\xi(\omega)+\eta_d\sqrt{\pi}\left(\delta(\omega+\omega_d)+\delta(\omega-\omega_d)\right)\right],
\end{align}
with the mechanical susceptibility $\epsilon_m(\omega)={i\omega}({\omega^2+2i\Gamma\omega-\nu^2})^{-1}$. The momentum correlation function is then given by the inverse transform $\braket{p(t)p(t')}=\frac{1}{2\pi}\int_{-\infty}^\infty d\omega\int_{-\infty}^\infty d\omega' e^{-i\omega t}e^{-i\omega' t'}\braket{p(\omega)p(\omega')}$ which for $\Gamma\ll\nu$ can be estimated as (averages over single noise terms vanish and $\epsilon_m(-\omega)=\epsilon_m^*(\omega)$):
\begin{align}
\braket{p(t)p(t')}=\frac{\eta_d^2}{2}\left[ e^{-(\Gamma+i\nu)(t-t')}+2|\epsilon_m(\omega_d)|^2\cos(\omega_d (t-t'))+\epsilon_m(\omega_d)^2 e^{-i\omega_d (t+t')}+\epsilon_m^* (\omega_d)^2 e^{i\omega_d (t+t')}\right].
\end{align}
Approximating $\epsilon_m(\omega_d)\approx\frac{1}{2}\frac{1}{\Gamma-i(\omega_d-\nu)}$ this can be expressed as
\begin{align}
\braket{p(t)p(t')}=\frac{1}{2}\left[e^{-(\Gamma+i\nu)(t-t')}+2|\beta|^2\cos\left(\omega_d(t-t')\right)+\beta\beta e^{-i\omega_d(t+t')}+\beta^*\beta^*e^{i\omega_d(t+t')} \right].
\end{align}
Furthermore, one can easily obtain
\begin{align}
\braket{p(t)}=\frac{1}{\sqrt{2\pi}}\int_{-\infty}^\infty d\omega e^{-i\omega t}\braket{p(\omega)}=\frac{1}{\sqrt{2}}\left(\beta^* e^{i\omega_d t}+\beta e^{-i\omega_d t}\right).
\end{align}

\section{Cavity dynamics}

\label{cavitydynamics}

We consider a driven IR-cavity with driving amplitude $\eta_d$ (frequency $\omega_d$) and a free space ``probe''-beam with amplitude $\eta_p$. The dynamics of the cavity-electron-vibron system is then encompassed in the following four equations for vibrational mode operators, cavity mode and polaron operator $\tilde{\sigma}$:
\begin{subequations}
\begin{align}
\dot{q}&=\nu p,\\
\dot{p}&=-2\Gamma p -\nu q -\sqrt{2}g(a^\dagger + a)+\xi (t),\\
\dot{a}&=-\left(\kappa+i\omega_c\right)a-i\sqrt{2}g q+\sqrt{2\kappa}a_{\text{in}}+\eta_d^c \cos(\omega_d t),\\
\dot{\tilde{\sigma}}&=-(\tilde{\gamma}-i\Delta_p)\tilde{\sigma}-2i\lambda g (a^\dagger+a)\tilde{\sigma}+\tilde{\Sigma}_{\text{in}},
\end{align}
\end{subequations}
Similar to the free space driving case, the coupling to the cavity leads to a periodic modulation of the electronic transition frequency $\tilde{\omega}_0=\omega_0+2\lambda g\left[a^\dagger(t)+a(t)\right]$. This can be approximated by eliminating the cavity mode (assuming $\omega_d\approx\omega_c$)
\begin{align}
\label{polaroncavity}
\dot{\tilde{\sigma}}&=-(\tilde{\gamma}-i\Delta_p)\tilde{\sigma}-2i\lambda g \frac{\eta_d^c}{\kappa} \cos(\omega_d t)\tilde{\sigma}+\tilde{\Sigma}_{\text{in}},
\end{align}
which is formally identical to the free space case in Eq.~(\ref{polaronfreespace}). For the calculation of the absorption spectrum, we again require the expectation value and correlation functions of the momentum quadrature which will be modified compared to the free space case due to coupling to the cavity. To this end, we will follow a standard optomechanical approach~\cite{aspelmeyer2014cavity}. In frequency space, the equations of motion read
\begin{subequations}
\begin{align}
-i\omega{q}(\omega)&=\nu p(\omega),\\
-i\omega{p}(\omega)&=-2\Gamma p(\omega) -\nu q(\omega) -\sqrt{2}g\left[a^\dagger(\omega) + a(\omega)\right]+\xi (\omega),\\
-i\omega{a}(\omega)&=-\left(\kappa+i\omega_c\right)a(\omega)-i\sqrt{2}g q(\omega)+\sqrt{2\kappa}a_{\text{in}}(\omega)+\sqrt{2\pi}\frac{\eta_d}{2}\left[ \delta(\omega-\omega_d)+\delta(\omega+\omega_d)\right],\\
-i\omega{a}^\dagger(\omega)&=-\left(\kappa-i\omega_c\right)a(\omega)+i\sqrt{2}g q(\omega)+\sqrt{2\kappa}a_{\text{in}}^\dagger(\omega)+\sqrt{2\pi}\frac{\eta_d}{2}\left[\delta(\omega+\omega_d)+\delta(\omega-\omega_d)\right].
\end{align}
\end{subequations}
Plugging the first equation into the second one, we can express
\begin{align}
p(\omega)=\epsilon_m\left(\omega\right)\left[\xi(\omega)-\sqrt{2}g\left(a^\dagger(\omega)+a(\omega)\right)\right]=:\epsilon_m(\omega)\left[\xi(\omega)+F_{\text{IR}}(\omega)\right],
\end{align}
with the force of the infrared field $F_{\text{IR}}(\omega)=-\sqrt{2}g\left[a^\dagger(\omega)+a(\omega)\right]$ and the mechanical susceptibility $\epsilon_m(\omega)=i\omega/(\omega^2+2i\Gamma\omega-\nu^2)$. The equations for the cavity field operators can be expressed as
\begin{subequations}
\begin{align}
a(\omega)&=\epsilon_c(\omega)\left[\sqrt{2\kappa}a_{\text{in}}-i\sqrt{2}gq+\sqrt{2\pi}\frac{\eta_d^c}{2}\left(\delta(\omega-\omega_d)+\delta(\omega+\omega_d)\right)\right],\\
a^\dagger(\omega)&=\epsilon_c^*(-\omega)\left[\sqrt{2\kappa}a_{\text{in}}^\dagger+i\sqrt{2}gq+\sqrt{2\pi}\frac{\eta_d^c}{2}\left(\delta(\omega+\omega_d)+\delta(\omega-\omega_d)\right)\right],
\end{align}
\end{subequations}
 with the susceptibility of the infrared cavity field $\epsilon_c(\omega)=\frac{1}{i(\omega_c-\omega)+\kappa}$.
 Defining the effective mechanical susceptibility $[\epsilon_m^{\text{eff}}]^{-1}(\omega)=\left[\epsilon_m^{-1}+2g^2\frac{\nu}{\omega}\left(\epsilon_c (\omega)-\epsilon_c^*(-\omega)\right)\right]$, we can express
\begin{align}
p(\omega)\!=\!\epsilon_m^{\text{eff}}(\omega)\!\left[\xi(\omega)\!-\!2\epsilon_c(\omega)\sqrt{\kappa}g a_{\text{in}}\!-\!2\epsilon_c^*(-\omega)\sqrt{\kappa}ga_{\text{in}}^\dagger-\epsilon_c(\omega)\sqrt{\pi}g\eta_d\left(\delta(\omega\!-\!\omega_d)\!+\!\delta(\omega\!+\!\omega_d)\right)\!-\!\epsilon_c^*(-\omega)\sqrt{\pi}g\eta_d\left(\delta(\omega\!+\!\omega_d)\!+\!\delta(\omega\!-\!\omega_d)\right)\right].
\end{align}
The effective mechanical susceptibility $\epsilon_m^{\text{eff}}$ can be expressed as:
\begin{align}
(\epsilon_m^{\text{eff}})^{-1}(\omega)&=\frac{\omega^2+2i\Gamma\omega-\nu^2}{i\omega}+2g^2\frac{\nu}{\omega}\left[\frac{\kappa}{\kappa^2+(\omega_c-\omega)^2}-\frac{\kappa}{\kappa^2+(\omega_c+\omega)^2}\right]\\\nonumber
&+2g^2\frac{\nu}{\omega}i\left[\frac{(\omega-\omega_c)}{\kappa^2+(\omega_c-\omega)^2}-\frac{(\omega+\omega_c)}{\kappa^2+(\omega_c+\omega)^2}\right].
\end{align}
From this, we can define an effective modified decay rate and frequency shift
\begin{subequations}
\begin{align}
\tilde{\Gamma}(\omega)&=\Gamma+g^2\frac{\nu}{\omega}\left[\frac{\kappa}{\kappa^2+(\omega_c-\omega)^2}-\frac{\kappa}{\kappa^2+(\omega_c+\omega)^2}\right],\\
\tilde{\nu}^2(\omega)&=\nu\left[\nu+2g^2\left(\frac{(\omega-\omega_c)}{\kappa^2+(\omega_c-\omega)^2}-\frac{(\omega+\omega_c)}{\kappa^2+(\omega_c+\omega)^2}\right)\right].
\end{align}
\end{subequations}
The absolute value squared of the effective mechanical susceptibility can then be written as
\begin{align}
|(\epsilon_m^{\text{eff}})(\omega)|^2=\frac{\omega^2}{(\tilde{\nu}^2(\omega)-\omega^2)^2+4\tilde{\Gamma}(\omega)^2\omega^2}.
\end{align}
 In the weak coupling regime $g<\kappa$, the effective mechanical susceptibility for $\omega_c=\nu$ becomes
\begin{align}
|(\epsilon_m^{\text{eff}})(\omega)|^2=\frac{\omega^2}{(\nu^2-\omega^2)^2+4\tilde{\Gamma}^2\omega^2},
\end{align}
so the only effect is the modification of the decay rate $\tilde{\Gamma}$.

\subsection{Calculation of momentum correlation function (weak coupling regime)}
For the calculation of the cavity-modified absorption spectrum one again requires the momentum correlation function. In frequency space, the momentum correlation function without the drive terms then becomes (using that all noise correlations except $\braket{\xi(\omega)\xi(\omega')}$ and $\braket{a_{\text{in}}(\omega)a_{\text{in}}^\dagger(\omega')}$ are zero):
\begin{align}
\label{momentumcorr}
\braket{p(\omega) p(\omega')}&=\epsilon_m ^{\text{eff}}(\omega)\epsilon_m^{\text{eff}}(\omega')[S_{\text{th}}(\omega)\delta(\omega+\omega')+\sir(\omega)\delta(\omega+\omega')].\\\nonumber
\end{align}
where we defined the infrared spectrum $\sir(\omega)=4\kappa g^2|\epsilon_c(\omega)|^2$. The momentum correlation function in time domain is then given by the Fourier transform
\begin{align}
\braket{p(t)p(t')}=\frac{1}{2\pi}\int_{-\infty}^{\infty}d\omega e^{-i\omega t}\int_{-\infty}^{\infty}d\omega' e^{-i\omega' t'}\braket{p(\omega)p(\omega')}.
\end{align}
Additionally to the two terms in Eq.~(\ref{momentumcorr}), there are 16 terms stemming from the driving of the cavity mode. For those terms we obtain (using that $\epsilon_m ^{\text{eff}}(-\omega)=\epsilon_m ^{\text{eff}}(\omega)^*$ and neglecting off-resonant terms $\epsilon_c(-\omega_d)$ and $\epsilon_c^*(-\omega_d)$):
\begin{align}
\braket{p(t)p(t')}_{\text{drive}}=\frac{g^2\eta_d^2}{2} \left[e^{-i\omega_d(t+t')}\epsilon_m ^{\text{eff}}(\omega_d)^2\epsilon_c(\omega_d)^2+2\cos(\omega_d(t-t'))|\epsilon_m ^{\text{eff}}(\omega_d)|^2|\epsilon_c(\omega_d)|^2+e^{i\omega_d(t+t')}\epsilon_m ^{\text{eff},*}(\omega_d)^2\epsilon_c^*(\omega_d)^2\right].
\end{align}
Similar to the free space driving, one finds that the momentum correlation function is not stationary. For the two terms independent of the drive, the first term describes the effective mechanical susceptibility weighted by the thermal spectrum:
\begin{align}
\braket{p(t)p(t')}_{\text{th}}=\frac{1}{2\pi}\int_{-\infty}^\infty d\omega e^{-i\omega(t-t')}|\epsilon_m ^{\text{eff}}(\omega)|^2 S_{\text{th}}(\omega)=\frac{1}{2\pi}\int_{-\infty}^\infty d\omega e^{-i\omega(t-t')}|\epsilon_m ^{\text{eff}}(\omega)|^2 \frac{2\Gamma\omega}{\nu}\left[\coth\left(\frac{\beta\omega}{2}\right)+1\right].
\end{align}
In the following we will focus on the case of zero temperature. The integral in this case picks only positive frequency components and can then be approximated by
\begin{align}
\braket{p(t)p(t')}_{\text{th}}=\frac{1}{2\pi}\int_{0}^\infty d\omega e^{-i\omega(t-t')}|\epsilon_m ^{\text{eff}}(\omega)|^2 \frac{4\Gamma\omega}{\nu}.
\end{align}
The second term describes the effect of the infrared field
\begin{align}
\braket{p(t)p(t')}_{\text{IR}}&=\frac{1}{2\pi}\int_{-\infty}^\infty d\omega e^{-i\omega(t-t')}|\epsilon_m ^{\text{eff}}(\omega)|^2 S_{\text{IR}}(\omega)=\frac{1}{2\pi}\int_{-\infty}^\infty d\omega e^{-i\omega(t-t')}|\epsilon_m ^{\text{eff}}(\omega)|^2\frac{4g^2\kappa}{\kappa^2+(\omega_c-\omega)^2}.
\end{align}
In the following we will calculate $\braket{p(t)p(t')}_{\text{IR}}$ and $\braket{p(t)p(t')}_{\text{th}}$ (for $\omega_c\approx\nu$). We will approximate the expressions under the integral with Lorentzians by expanding around the resonances and only keeping the leading order terms of the expansion. In the weak coupling regime, we can expand around $\omega=\nu+\delta$:
\begin{align}
\braket{p(t)p(t')}_{\text{th}}&=\frac{1}{2\pi}e^{-i\nu(t-t')}\int_{-\infty}^\infty d\delta e^{-i\delta(t-t')}\frac{\Gamma}{\tilde{\Gamma}}\frac{\tilde{\Gamma}}{\delta^2+\tilde{\Gamma}^2}=\\\nonumber
&=\frac{\Gamma}{2\tilde{\Gamma}}e^{-(\tilde{\Gamma}+i\nu)(t-t')}.
\end{align}

\begin{align}
\braket{p(t)p(t')}_{\text{IR}}&=\frac{1}{2\pi}e^{-i\nu(t-t')}\int_{-\infty}^\infty d\delta e^{-i\delta(t-t')}\frac{g^2}{\kappa\tilde{\Gamma}}\frac{\tilde{\Gamma}}{\delta^2+\tilde{\Gamma}^2}=\\\nonumber
&=C\frac{\Gamma}{2\tilde{\Gamma}}e^{-(\tilde{\Gamma}+i\nu)(t-t')},
\end{align}
with $C=\frac{g^2}{\kappa\Gamma}$. All together this just gives $\frac{\Gamma}{2\tilde{\Gamma}}(1+C)e^{-(\tilde{\Gamma}+i\nu)(t-t')}=\frac{1}{2}e^{-(\tilde{\Gamma}+i\nu)(t-t')}$, i.e., the only effect is the modification of the spectral linewidth of $\Gamma$. For $\nu\neq\omega_c$, the generalized cooperativity can be expressed as as $C(\omega)=\frac{g^2\kappa/\Gamma}{\kappa^2+(\omega-\omega_c)^2}$. The total momentum correlation function is then given by
\begin{align}
\braket{p(t)p(t')}=\braket{p(t)p(t')}_{\text{drive}}+\braket{p(t)p(t')}_{\text{th}}+\braket{p(t)p(t')}_{\text{IR}}.
\end{align}

\subsection{Cavity-modified absorption spectrum}

The cavity-modified ground-excited electronic coherence is given by a formal integration of Eq.~(\ref{polaroncavity})
\begin{align}
\braket{\sigma(t)}=\eta_p\int_{-\infty}^t dt' e^{-(\gamma-i\Delta_p)(t-t')} e^{-2i\lambda g\eta_d\left[\sin(\omega_d t)-\sin(\omega_d t')\right]/(\omega_d\kappa)}\braket{\mathcal{D}(t)\mathcal{D}^\dagger (t')}.
\end{align}
The steady-state population can then be calculated analogously to Appendix \ref{freespacedriving} which yields
\begin{align}
\braket{\sigma^\dagger\sigma}=\frac{\eta_p^2}{\gamma}\!\sum_{n=0}^\infty\sum_{m,\ell=-\infty}^\infty\frac{\lambda^{2n}}{n!}\frac{f_{\text{FC}}(\tilde{\gamma}+\!n\tilde{\Gamma})\mathcal{J}_m\!\left(\frac{2\lambda g\eta_d^c}{\kappa\omega_d}\right)^2\BesselJ_\ell\left(2\lambda|\beta_{\text{c}}|\right)^2}{(\tilde{\gamma}+\!n\tilde{\Gamma})^2+(\Delta_p\!-\!n\nu-\!(m+\ell)\omega_d)^2},
\end{align}
with the cavity-modified vibrational occupation $|\beta_{\text{c}}|^2= g^2\eta_d^2|\epsilon_m^{\text{eff}}(\omega_d)|^2|\epsilon_c (\omega_d)^2|$. For resonant driving of the cavity $\omega_d=\omega_c$, this can be approximated by $|\beta_{\text{c}}|^2=\frac{g^2}{\kappa^2}\frac{(\eta_d^c/2)^2}{\tilde{\Gamma}^2+(\omega_d-\nu)^2}$.

\end{document}